\documentclass[preprint,12pt]{elsarticle}

\usepackage{amssymb}

\journal{Pattern Recognition}
\usepackage{algorithm}
\usepackage{algorithmic}
\usepackage{multirow}
\usepackage{booktabs}
\usepackage{arydshln}
\usepackage{hyperref}
\usepackage{enumitem}
\usepackage{amsmath}

\begin{document}

\begin{frontmatter}



\title{High-level Codes and Fine-grained Weights for Online Multi-modal Hashing Retrieval}

\author[label1]{Yu-Wei Zhan\corref{Eqauthor}}

\author[label2]{Xiao-Ming Wu\corref{Eqauthor}}

\author[label1,label3]{Xin Luo\corref{mycorrespondingauthor}}

\cortext[Eqauthor]{Equal contribution}

\cortext[mycorrespondingauthor]{Corresponding author (Email: luoxin.lxin@gmail.com)}

\author[label5]{Yinwei Wei}
\author[label1]{Xin-Shun Xu}

\affiliation[label1]{organization={Shandong University},
            city={Jinan},
            country={China}}

\affiliation[label2]{organization={Sun Yat-sen University},
            city={Guangdong},
            country={China}}

\affiliation[label3]{organization={Yunnan Key Laboratory of Software Engineering},
            city={Yunnan},
            country={China}}
            
\affiliation[label5]{organization={Monash University},
            city={Melbourne},
            country={Australia}}



\begin{abstract}
In the real world, multi-modal data often appears in a streaming fashion, and there is a growing demand for similarity retrieval from such non-stationary data, especially at a large scale. In response to this need, online multi-modal hashing has gained significant attention. However, existing online multi-modal hashing methods face challenges related to the inconsistency of hash codes during long-term learning and inefficient fusion of different modalities. In this paper, we present a novel approach to supervised online multi-modal hashing, called \textbf{H}igh-level \textbf{C}odes, \textbf{F}ine-grained \textbf{W}eights (\textbf{HCFW}). To address these problems, HCFW is designed by its non-trivial contributions from two primary dimensions:
1) Online Hashing Perspective. To ensure the long-term consistency of hash codes, especially in incremental learning scenarios, HCFW learns high-level codes derived from category-level semantics. Besides, these codes are adept at handling the category-incremental challenge. 
2) Multi-modal Hashing Aspect. HCFW introduces the concept of fine-grained weights designed to facilitate the seamless fusion of complementary multi-modal data, thereby generating multi-modal weights at the instance level and enhancing the overall hashing performance.
A comprehensive battery of experiments conducted on two benchmark datasets convincingly underscores the effectiveness and efficiency of HCFW.
\end{abstract}



\begin{keyword}


Multi-modal Retrieval; Online Hashing; Learning to Hash
\end{keyword}

\end{frontmatter}


\section{Introduction}
The rapid advancement of hardware and application software has given rise to an escalating demand for similarity search mechanisms, particularly in dealing with extensive multimedia datasets. In response to this need, hashing \cite{wang2010sequential} has emerged as one of the most prominent techniques for approximate nearest neighbor searches, due to the high retrieval speed and low storage cost. 
Based on the availability of supervisory information, hashing methods can be broadly categorized into supervised \cite{wang2021high, liu2014discrete, shrivastava2015asymmetric} and unsupervised approaches \cite{norouzi2011minimal, doan2020efficient, zhang2014supervised, shen2015supervised}.

In recent years, there has been a growing demand for efficient retrieval from streaming sources in various real-world applications. However, traditional hashing methods are primarily designed for batch-based learning. When new data arrives to train such batch-based hashing models, a cumbersome process of accumulating both the new and old data is necessitated for model retraining. This approach presents practical challenges in real-world deployments, as frequent retraining incurs significant computational costs and the storage requirements for all data can be prohibitively large. To overcome such limitations, a series of methods termed online hashing\cite{li2022recent} have been proposed and achieved satisfactory performance. More specifically, existing online hashing methods can be classified into three categories, i.e., uni-modal \cite{weng2019fast,tu2021partial,lin2020hadamard,weng2021online,jin2021asynchronous,li2021online,TianNWK21}, cross-modal \cite{qi2017online,zhan2022discrete}, and multi-modal \cite{xie2017dynamic,lu2019flexible, wu2021online}. 
Uni-modal methods are specialized in retrieving data within the same modality, facilitating intra-modal searches. Cross-modal techniques are engineered to support queries spanning multiple modalities, enabling inter-modal inquiry. 
Our research focuses on online multi-modal hashing. This domain diverges significantly from the uni-modal and cross-modal settings, where a single modality suffices for querying. Instead, online multi-modal hashing needs the incorporation of multi-modal features, accommodating data with multiple modalities within the retrieval process.

To the best of our knowledge, research into the intricate domain of online multi-modal hashing has been explored by only three notable works: Online Dynamic Multi-View Hashing (ODMVH) \cite{xie2017dynamic}, Flexible Online Multi-modal Hashing (FOMH) \cite{lu2019flexible}, and Online enhAnced SemantIc haShing (OASIS) \cite{wu2021online}. Despite their commendable performance, these methods still have certain limitations. 1) Notably, ODMVH pioneered the field of online multi-modal hashing as the inaugural method. It is conceived as an unsupervised approach. As widely recognized in hashing literature, neglecting the supervised information may lead to insufficient hash learning and poor accuracy when such information is available. 2) FOMH is a supervised approach that attains state-of-the-art performance. Nonetheless, it falls short when confronted with the category incremental problem associated with streaming data. FOMH implicitly assumes that all categories are present in the initial data chunk and doesn't account for the possibility of new, unseen categories emerging in subsequent chunks. 3) OASIS has emerged as a pioneer in addressing the critical challenge of category incremental learning in the field of online multi-modal hashing. However, it inadvertently overlooks the long-term consistency of hash codes in online learning, i.e., category compactness and semantic relevance, and it fails to explicitly fuse complementary multi-modal information. 

To address the aforementioned challenges, this paper introduces an innovative method named High-level Codes, Fine-grained Weights (abbreviated as HCFW). HCFW presents a novel solution for the relatively unexplored domain of online multi-modal hashing. \textbf{Online Learning Enhancement:} In the context of incremental learning, particularly in class-incremental scenarios, HCFW learns high-level codes derived from category-level semantics to ensure matrix dimension matching and the consistency of hash codes across multiple rounds of learning. These high-level codes guarantee that hash codes of instances from the same category in different rounds are compact, and hash codes of semantically similar categories in different rounds are similar. In addition, the well-designed matrix dimension match is able to handle the category incremental problem. \textbf{Multi-Modal Fusion:} Regarding multi-modal hashing, HCFW introduces the concept of fine-grained weights at the instance level to achieve a more refined fusion of modalities. This innovative approach enables the effective fusion of complementary multi-modal data, resulting in the generation of high-quality hash codes. The main contributions of our model are summarized as follows:
\begin{itemize}
\item A new online multi-model hashing method called HCFW is proposed. HCFW attempts to learn high-level hash codes from semantics and construct good relevance about categories to ensure matrix dimension matching and long-term consistency in online learning. Additionally, HCFW adeptly addresses the category-incremental problem through the strategic utilization of high-level codes.
\item HCFW introduces fine-grained weights at the instance level to facilitate the effective fusion of multi-modal features during hash code generation. Different samples have their individual modality fusion weights. 
\item Extensive experiments conducted on two benchmark datasets underscore the exceptional performance of our method in terms of both accuracy and computational efficiency. Besides, We will release the code for HCFW soon and hope that it could facilitate other researchers and the community.
\end{itemize}

\section{Revisiting Online Multi-Modal Hashing}
\subsection{Multi-Modal Hashing}
Given the distinction in data, existing hashing methods can be broadly categorized into three primary types: uni-modal hashing \cite{gong2012iterative,liu2019supervised2,9712384}, cross-modal hashing \cite{yang2020nonlinear,zhou2014latent,sun2019supervised,wang2020online}, and multi-modal hashing \cite{lu2019online,zhu2020flexible,zhang2011composite}.
As mentioned earlier, \textbf{multi-modal hashing}, which is also called multi-view hashing, tries to combine heterogeneous multi-modal features during both training and querying. Such a setting is different from uni-modal and cross-modal hashing where only one modality is provided in the period of querying. 

Representative multi-modal hashing methods include but are not limited to Composite Hashing with Multiple Information Sources (CHMIS) \cite{zhang2011composite}, Multiple Feature Hashing (MFH) \cite{song2011multiple}, Multi-view Anchor Graph Hashing (MVAGH) \cite{kim2013multi}, Multi-view Alignment Hashing (MAH) \cite{liu2015multiview} Multi-View Latent Hashing (MVLH) \cite{shen2015multi},  Multiview Discrete Hashing (MVDH) \cite{shen2018multiview}, Compact Kernel Hashing with Multiple Features (MFKH)  \cite{liu2012compact}, and Discrete Multi-View Hashing (DMVH) \cite{yang2017discrete}. However, most existing multi-modal hashing methods are batch-based, assuming that all data is needed for model training. However, this assumption is impractical for real-world applications that often deal with streaming data. Traditional batch-based models must accumulate all previous data and retrain when new, unseen data arrives. This approach is inflexible, inefficient, and resource-intensive, especially with larger data volumes.

\subsection{Online Hashing}
In real-world applications, data always comes in a streaming fashion, and conducting similarity retrieval among non-stationary data is one of the most important research scopes. Due to the high retrieval speed and low storage cost, online hashing\cite{li2022recent} has become one promising solution to fulfill the above purpose. Different from batch-based hashing, \textbf{online hashing} is born for online retrieval tasks and is capable of efficiently learning from streaming data as they could be updated only based on the newly coming data while preserving knowledge learned from old seen data.

\begin{figure}[t]
\centering
\includegraphics[width=1\columnwidth]{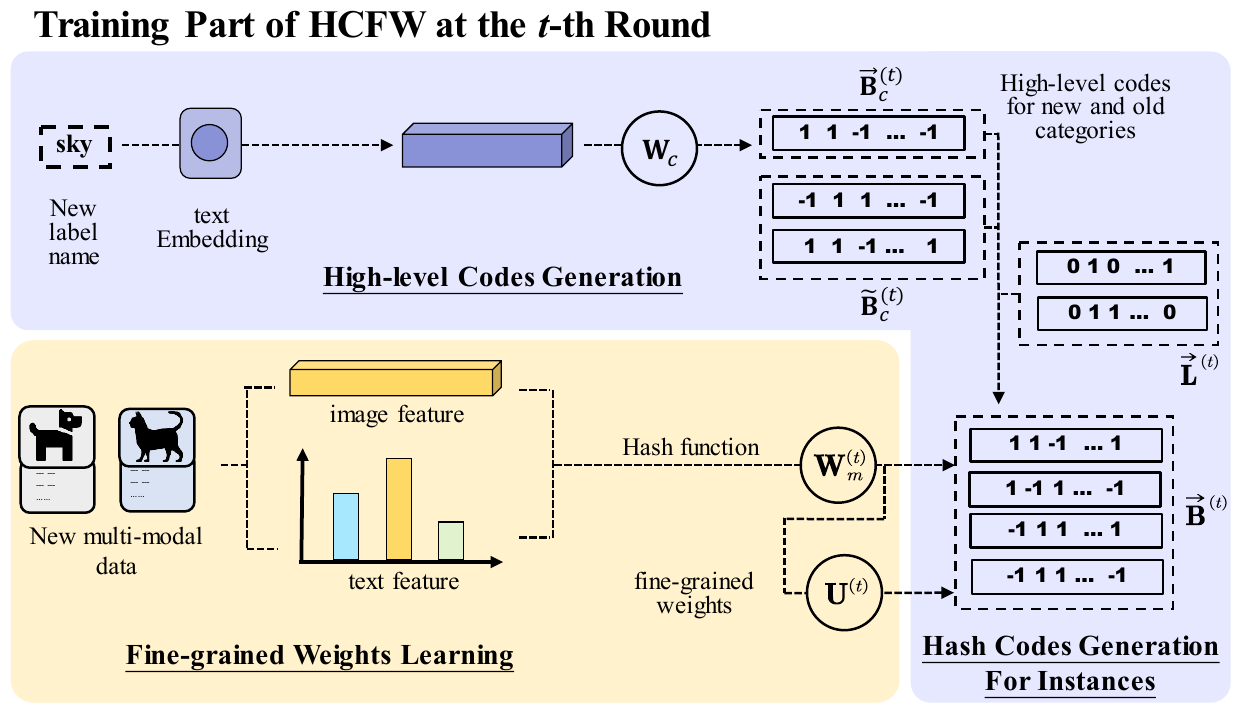}
\caption{The framework of the proposed HCFW. Without loss of generality, the training procedure of HCFW at the t-th round is illustrated. It contains two parts, i.e., high-level code generation and fine-grained weights. }
\label{model_figure}
\vspace{-0.2cm}
\end{figure}

More specifically, existing online hashing methods can be classified into three categories, i.e., uni-modal \cite{weng2019fast,tian2021concept,lin2020hadamard,weng2021online,jin2021asynchronous,li2021online,TianNWK21}, cross-modal \cite{qi2017online,chen2019extensible,zhan2022discrete}, and multi-modal \cite{xie2017dynamic,lu2019flexible}. Online uni-modal hashing adopts queries from one modality to search for similar instances within the same modality \cite{cakir2017online,wu2019deep,chen2019making,tian2020complementary,zhan2021weakly}. Representative methods include but not limited to Online Hashing \cite{huang2013online}, Online Sketching Hashing \cite{leng2015online}, Adaptive Hashing \cite{cakir2015adaptive}, Incremental Hashing \cite{ng2017incremental}, Online Hashing with Mutual Information \cite{cakir2017mihash}, FasteR Online Sketching Hashing \cite{chen2017frosh}, Hadamard Codebook based Online Hashing \cite{lin2018supervised}, and Online Hashing with Efficient Updating \cite{weng2020online}. Online cross-modal hashing supports cross-modal retrieval tasks, e.g., using texts as queries to retrieve similar images. Online Cross-modal Hashing \cite{xie2016online}, GrowBit \cite{mandal2018growbit}, Online Latent Semantic Hashing \cite{yao2019online}, Label Embedding Online Hashing \cite{wang2020label}, Online Collective Matrix Factorization Hashing \cite{wang2020online}, and Online Label Consistent Hashing \cite{yi2021efficient} belong to this category. Although impressive performance has been obtained by the aforementioned online hashing methods, very little effort has been put into the \textbf{online multi-modal hashing}. To the best of our knowledge, only three already published works attempt to investigate online multi-modal hashing: Online Dynamic Multi-View Hashing (ODMVH) \cite{xie2017dynamic}, Flexible Online Multi-modal Hashing (FOMH) \cite{lu2019flexible}, and Online enhAnced SemantIc haShing (OASIS) \cite{wu2021online}. As analyzed above, they still suffer from some limitations e.g., overlook the long-term consistency of hash codes and the explicit fusion of complementary multi-modal information.

\section{HCFW}
Our approach adopts a two-step hashing strategy, breaking down the learning process into hash code learning and hash function learning. Accordingly, the proposed HCFW comprises two primary components: high-level code generation during hash code learning and fine-grained weight acquisition in hash function learning. 
The overall architecture of HCFW is illustrated in Figure \ref{model_figure}.

\subsection{Problem Definition and Notations}
\subsubsection{Problem Definition}
This paper addresses the challenge of online multi-modal hashing, which can be formally defined as follows: 1) The training data arrives in a streaming fashion and comprises multiple modalities. In such non-stationary environments, the category incremental problem may arise, where newly arriving data may introduce previously unseen categories. 2) After receiving a chunk of data (referred to as the current data round), multi-modal hashing methods promptly learn their hash codes. Once the hash codes in the current round are obtained, this data may not be utilized for training in subsequent data rounds. 3) To encode out-of-sample multi-modal queries into binary codes, the hash function must be learned based on the streaming training data.

\subsubsection{Notations}
In this paper, following existing multi-modal hashing literature \cite{xie2017dynamic,lu2019flexible} and for the sake of clear representation, we present our model based on image and text modalities and stipulate that image is the first modality and text is the second one. If there are more modalities, the processing paradigm is the same.

As data comes in a streaming fashion, we thus detail our model in the context of the $t$-th round (the $t$-th data chunk) without loss of generality, where there comes $n^{(t)}$ instances and has $N^{(t)}$ already seen instances (observed before current round and $N^{(t)}= n^{(1)}+ \cdots+n^{(t-1)} $). We define $\overrightarrow{\mathbf{X}}_{m}^{\left( t \right)}\in \mathbb{R} ^{d_m\times n^{\left( t \right)}}$ as the newly arriving data of the $m$-th modality at the $t$-th round and $\widetilde{\mathbf{X}}_{m}^{\left( t \right)}\in \mathbb{R} ^{d_m\times N^{\left( t \right)}}$ as the previously arrived data of the $m$-th modality at the $t$-th round, where $d_m$ is the dimension of  modality features. Moreover, we denote $\overrightarrow{\mathbf{B}}^{( t )}\in \{ -1,1 \} ^{r\times n^{( t )}}$ and $\widetilde{\mathbf{B}}^{\left( t \right)}\in \left\{ -1,1 \right\} ^{r\times N^{\left( t \right)}}$ as hash codes of the new data and previous arrived data at the $t$-th round, where $r$ is the length of hash codes. Similarly, we define $\overrightarrow{\mathbf{L}}^{\left( t \right)}\in \left\{ 0,1 \right\} ^{\left( c_{n}^{\left( t \right)}+c_{o}^{\left( t \right)} \right) \times n^{\left( t \right)}}$ and $\widetilde{\mathbf{L}}^{\left( t \right)}\in \left\{ 0,1 \right\} ^{c_{o}^{\left( t \right)}\times N^{\left( t \right)}}$ as labels of new data and old data at round $t$, where $c_{o}^{\left( t \right)}$ and $c_{n}^{\left( t \right)}$  are the number of old categories and newly arriving categories at the $t$-th round.

\subsection{High-Level Codes}
\subsubsection{Category Incremental Problem}
For non-stationary data, the class-incremental problem poses a significant challenge, yet there is little literature on online multi-modal hashing addressing this issue. Some of the difficulties hindering research in this area include: 1) Dimension mismatch caused by differences in the dimensions of label matrices. 2) The issue of inconsistent hash codes across multiple rounds of learning. Existing methods often fail to maintain consistency and compactness of hash codes during long-term learning processes, leading to the loss of previously acquired information.

$ \| r\mathbf{S}- \mathbf{B}^{T}\mathbf{B} \| _{F}^{2} $ and $ \| \mathbf{L}- \mathbf{Q}\mathbf{B} \| _{F}^{2} $ are the most commonly used terms to embed semantic information, where $\mathbf{S}={\mathbf{L}}^{( t )}{\mathbf{L}}^{( t )T}$ is pairwise similarity of data and $\mathbf{Q}$ is projection between hash codes and labels. 
The first challenge arises from the introduction of these terms, which may not be suitable for the class-incremental problem due to dimension mismatch. For example, OASIS \cite{wu2021online} has pointed out that some existing methods are limited when computing pairwise similarity between new data and old data, i.e., $\widetilde{\mathbf{L}}^{( t )}\overrightarrow{\mathbf{L}}^{( t )T}$. Besides, some online hashing methods \cite{su2020online,wang2020label} adapt it into $\| \overrightarrow{\mathbf{L}}^{( t )}-\mathbf{Q}\overrightarrow{\mathbf{B}}^{( t )} \| _{F}^{2}+\| \widetilde{\mathbf{L}}^{( t )}-\mathbf{Q}\widetilde{\mathbf{B}}^{( t )} \| _{F}^{2}$ and fail to handle the $c_{n}^{( t )}\neq 0$ situation. For another instance, in \cite{yao2019online}, the usage of supervised information can be abstracted as $\widetilde{\mathbf{L}}^{( t )} \widetilde{\mathbf{L}}^{(t) T}- \overrightarrow{\mathbf{L}}^{( t )} \overrightarrow{\mathbf{L}}^{(t) T}$ which may face dimensions mismatching when $c_{n}^{( t )}\neq 0$ . To overcome those adversities, we propose a new strategy termed high-level codes to deal with it. The details are shown in Section \ref{High-Level Codes Generation}.

Compared to the first challenge, the second issue concerning inconsistent hash codes is more fundamental and critical. Batch-based methods generate hash codes for all data simultaneously, ensuring uniformity across all codes. In contrast, online-based methods, during their extended learning processes, inevitably encounter inconsistencies in hash codes, particularly for instances belonging to the same or similar categories. These methods lack a mechanism to explicitly ensure that newly generated hash codes remain consistent with those previously stored in the database. For example, it's difficult to guarantee the compactness of hash codes from the same class and the similarity of hash codes from similar classes across multiple rounds. As more rounds progress, the problem of inconsistent hash codes becomes increasingly apparent, and the introduction of unseen classes further complicates the issue.


\subsubsection{High-Level Codes Generation }\label{High-Level Codes Generation}
Initially, we learn hash codes for each category instead of each instance, termed as \textbf{high-level codes}, and then leverage them to generate hash codes for instances. This method allows us to directly learn hash codes for new categories in the current round, avoiding the need for label matrices during learning and overcoming the dimensionality mismatch issue. Additionally, we ensure that the previously learned high-level codes (hash codes for old categories) remain unchanged when learning hash codes for new categories, and we do not modify the high-level codes for all categories when learning hash codes for instances. This approach effectively prevents information loss and preserves hash code consistency across rounds. Further details are provided below.
 
Firstly, to capture the semantic information, we employ the text embedding method to generate the semantic vectors for new categories at the $t$-th round, which is,
\begin{equation}
\begin{aligned}\label{high-level semantics}
\overrightarrow{\mathbf{K}}_{j}^{\left( t \right)}=\mathrm{Text\_Embedding}\left( \overrightarrow{\mathbf{Y}}_{j}^{\left( t \right)} \right) , j=1,2...c_{n}^{\left( t \right)},
\end{aligned}
\end{equation}
where  $ \overrightarrow{\mathbf{Y}}_{j}^{\left( t \right)} (j=1,2...c_{n}^{\left( t \right)})$ represents the category name (like "tree" and "sky") of $c_{n}^{\left( t \right)}$ new categories first appearing at round $t$,  $\overrightarrow{\mathbf{K}}_{j}^{\left( t \right)}$
represents the semantic vector of the $j$-th category. Then we can acquire the semantics matrix of new categories at the $t$-th round
$
\overrightarrow{\mathbf{K}}^{( t )}=[ \overrightarrow{\mathbf{K}}_{1}^{( t )}\,\,\overrightarrow{\mathbf{K}}_{2}^{( t )}\,\,... \overrightarrow{\mathbf{K}}_{c_{n}^{( t )}}^{\left( t \right)} ] \in \mathbb{R} ^{k\times c_{n}^{( t )}}
$
where $k$ is the dimensionality of the word2vec vector. It is worth noting that  
$
\widetilde{\mathbf{K}}^{( t )}=[ \widetilde{\mathbf{K}}_{1}^{( t )}\,\,\widetilde{\mathbf{K}}_{2}^{( t )}\,\,... \widetilde{\mathbf{K}}_{c_{o}^{( t )}}^{( t )} ] \in \mathbb{R} ^{k\times c_{o}^{( t )}}
$
of old categories is calculated at previous rounds and can be used directly at round $t$.

After generating semantics, we straightforwardly embed the semantics $\widetilde{\mathbf{K}}^{( t )} $ and $\overrightarrow{\mathbf{K}}^{( t )}$ into hash codes to learn high-level codes. More specifically, we hope category-level hash codes may reconstruct the high-level semantic matrix and the loss function is formulated as,
\begin{equation}
\begin{aligned}\label{Category Incremental Learning}
\underset{\overrightarrow{\mathbf{B}}_{c}^{\left( t \right)}, \mathbf{W}_c}{\min}&\left\| \overrightarrow{\mathbf{K}}^{\left( t \right)}-\mathbf{W}_{c}^{T}\overrightarrow{\mathbf{B}}_{c}^{\left( t \right)} \right\| _{F}^{2}+\left\| \widetilde{\mathbf{K}}^{\left( t \right)}-\mathbf{W}_{c}^{T}\widetilde{\mathbf{B}}_{c}^{\left( t \right)} \right\| _{F}^{2}
\\&
s.t. \overrightarrow{\mathbf{B}}_{c}^{\left( t \right)}\in \left\{ -1,1 \right\} ^{r\times c_n^{(t)}}.
\end{aligned}
\end{equation}
where $\overrightarrow{\mathbf{B}}_{c}^{\left( t \right)}$ and $\widetilde{\mathbf{B}}_{c}^{\left( t \right)}$ are the category-level hash code matrix of new and old categories, which termed \textbf{high-level codes} in our paper, $\mathbf{W}_c$ is the transformation matrix. It can be seen from the above formula that if there is no new category at the current round, there is no need to carry out this learning process. Although both our method and Hadamard matrix-based strategy could construct relevance among categories, ours could further embed the semantics into model learning which is more explainable. In Section \ref{model_design}, comparisons between those two strategies are shown.

\subsubsection{Hash Codes Generation for Instances}
Based on the high-level codes of categories, we generate the hash codes for instances,
\begin{equation}\label{B}
\begin{aligned}
\overrightarrow{\mathbf{B}}^{\left( t \right)}=\mathrm{sign}\left( \left[ \widetilde{\mathbf{B}}_{c}^{\left( t \right)}\ \ \overrightarrow{\mathbf{B}}_{c}^{\left( t \right)} \right] \overrightarrow{\mathbf{L}}^{\left( t \right)} \right),
\end{aligned}
\end{equation}
where $\mathrm{sign}( \cdot ) $ is the sign function, $\widetilde{\mathbf{B}}_{c}^{\left( t \right)} \in \mathbb{R} ^{r\times c_{o}^{( t )}}$, $\overrightarrow{\mathbf{B}}_{c}^{\left( t \right)} \in \mathbb{R} ^{r\times c_{n}^{( t )}}$, and $\overrightarrow{\mathbf{L}}^{\left( t \right)}\in \left\{ 0,1 \right\} ^{\left( c_{n}^{\left( t \right)}+c_{o}^{\left( t \right)} \right) \times n^{\left( t \right)}}$. The above function means that the hash codes of the instance are the linear combination of the high-level hash codes of the categories the instance belongs. Here, we rely on the label matrix of new instances just to generate the hash codes of them, thus the dimension mismatching problem caused by learning using both new and old label matrices in previous papers would not occur.


To address the category incremental problem, we employ Eqn. (\ref{Category Incremental Learning}). With the high-level codes, we bypass the direct usage of labels and explicitly model the new categories, which helps to deal with the first difficulty, i.e., the dimensionality mismatch problem. Furthermore, with the help of the transformation matrix $\mathbf{W}_c$ and high-level codes that have been learned and won't change again, knowledge from both old and new data could interact with each other and alleviate the inconsistent hash codes problem. 


\subsubsection{Optimization}
To solve the problem in Eqn.(\ref{Category Incremental Learning}), we introduce an alternating optimization algorithm that owns several iterations. Each iteration contains two steps, i.e., $\mathbf{W}_c$-step and $\overrightarrow{\mathbf{B}}_{c}^{\left( t \right)}$-step. In each step, we optimize one variable with the other fixed. 

\textbf{$\mathbf{W}_c$-step}. With $\overrightarrow{\mathbf{B}}_{c}^{\left( t \right)}$ fixed, we could directly take the derivative of Eqn.(\ref{Category Incremental Learning}) w.r.t. $\mathbf{W}_c$ to zero and the solution for $\mathbf{W}_c$ can be obtained as shown below,
\begin{equation}
\begin{aligned}\label{W_c}
\mathbf{W}_c=\left( \overrightarrow{\mathbf{B}}_{c}^{\left( t \right)}\overrightarrow{\mathbf{B}}_{c}^{\left( t \right) T}+\widetilde{\mathbf{B}}_{c}^{\left( t \right)}\widetilde{\mathbf{B}}_{c}^{\left( t \right) T} \right) ^{-1}\left( \overrightarrow{\mathbf{B}}_{c}^{\left( t \right)}\overrightarrow{\mathbf{K}}^{\left( t \right) T}+
\right.\left.\widetilde{\mathbf{B}}_{c}^{\left( t \right)}\widetilde{\mathbf{K}}^{\left( t \right) T} \right).
\end{aligned}
\end{equation}

\textbf{$\overrightarrow{\mathbf{B}}_{c}^{\left( t \right)}$-step}. For $\overrightarrow{\mathbf{B}}_{c}^{\left( t \right)}$, we discretely update it row by row, i.e., learning one row of $\overrightarrow{\mathbf{B}}_{c}^{\left( t \right)}$ each time with other rows fixed. Without loss of generality, we take the $j$-th row as an example. We define $\overrightarrow{\mathbf{B}}_{cj}^{\left( t \right)}$, $\mathbf{W}_{cj}$ as the transpose of the $j$-th row of $\overrightarrow{\mathbf{B}}_{c}^{\left( t \right)}$, $\mathbf{W}_c$, and $\overrightarrow{\mathbf{B'}}_{c}^{\left( t \right)}$, $\mathbf{W'}_c$ as the remaining part of $\overrightarrow{\mathbf{B}}_{c}^{\left( t \right)}$, $\mathbf{W}_c$ after removing the $j$-th row. By fixing $\mathbf{W}_c$, the problem to update the $j$-th row of $\overrightarrow{\mathbf{B}}_{c}^{\left( t \right)}$ can be simplified as,
\begin{equation}
\begin{aligned}
&\underset{\overrightarrow{\mathbf{B}}_{c}^{\left( t \right)},}{\min}\left\| \overrightarrow{\mathbf{K}}^{\left( t \right)}-\mathbf{W}_{c}^{T}\overrightarrow{\mathbf{B}}_{c}^{\left( t \right)} \right\| _{F}^{2}
\\
&=\left\| \mathbf{W}_{cj}\overrightarrow{\mathbf{B}}_{cj}^{\left( t \right) T}+\mathbf{W'}_{c}^{T}\overrightarrow{\mathbf{B'}}_{c}^{\left( t \right)} \right\| _{F}^{2}-2\mathrm{tr}\left( \mathbf{Q}_j\overrightarrow{\mathbf{B}}_{cj}^{\left( t \right) T} \right) +\mathrm{const}
\\
&=2\mathrm{tr}\left( \overrightarrow{\mathbf{B}}_{cj}^{\left( t \right) T}\left( \overrightarrow{\mathbf{B'}}_{c}^{\left( t \right) T}\mathbf{W'}_c\mathbf{W}_{cj}-\mathbf{Q}_j \right) \right) +\mathrm{const},
\end{aligned}
\end{equation}
where $\mathbf{Q}=\mathbf{W}_c\overrightarrow{\mathbf{K}}^{\left( t \right)}$ and $\mathbf{Q}_j$ is the transpose of the $j$-th row of $\mathbf{Q}$. Then, we can easily get the updating formula for $\overrightarrow{\mathbf{B}}_{cj}^{\left( t \right)}$ as,
\begin{equation}
\begin{aligned}\label{B_cj}
\overrightarrow{\mathbf{B}}_{cj}^{\left( t \right)}=\mathrm{sign}\left( \mathbf{Q}_j-\overrightarrow{\mathbf{B'}}_{c}^{\left( t \right) T}\mathbf{W'}_c\mathbf{W}_{cj} \right). 
\end{aligned}
\end{equation}
By repeating the above steps, we could finally optimize Eqn.(\ref{Category Incremental Learning}).

\subsection{Fine-Grained Weights}
\subsubsection{Hash Function Learning}
In hashing literature, once we obtained the hash codes of training instances, the hash functions could be learned through the two-step hashing strategy \cite{lin2013general}. Its effectiveness is adequately corroborated in both batch-based hashing \cite{lin2015semantics,kang2016column} and online hashing \cite{lin2018supervised,wang2020label}. Although more complex hash functions could be designed \cite{lin2013general,lin2015semantics,kang2016column}, e.g., using neural networks, the linear regression based function is adopted in HCFW due to its efficiency and satisfactory performance for online hashing scenarios. We formalize the objective function for hash function learning as follows,
\begin{equation}\label{Hash Function Loss}
\begin{aligned}
&\underset {\mathbf{W}_{m}^{( t )}}{\min}\sum_{m=1}^2{( \| \overrightarrow{\mathbf{B}}^{( t )}-\mathbf{W}_{m}^{( t )}  \overrightarrow{\mathbf{X}}_{m}^{( t )}  \| _{F}^{2}+\| \widetilde{\mathbf{B}}^{( t )}-}\mathbf{W}_{m}^{( t )} \widetilde{\mathbf{X}}_{m}^{( t )}  \| _{F}^{2}  +\theta \| \mathbf{W}_{m}^{( t )} \| _{F}^{2} ),
\end{aligned}
\end{equation}
where $\mathbf{W}_{m}^{( t )}$ is the hash projection for modality $m$ and $\theta$ balances the regularization term.

To solve Eqn.(\ref{Hash Function Loss}), each modality $\mathbf{W}_{m}^{( t )}$ can be optimized independently. Without loss of generality, we take the $m$-th modality as an example. The closed-form solution is obtained by setting the derivative of Eqn.(\ref{Hash Function Loss})  w.r.t. $\mathbf{W}_{m}^{( t )}$ to zero,
\begin{equation}\label{W}
\begin{aligned}
&\mathbf{W}_{m}^{( t )}=\mathbf{D}_{1}^{( t )}( \mathbf{D}_{2}^{( t )}+\theta \mathbf{I} ) ^{-1},
\end{aligned}
\end{equation}
where
\begin{equation}
\begin{aligned}
&\mathbf{D}_{1}^{( t )}=\overrightarrow{\mathbf{B}}^{( t )} \overrightarrow{\mathbf{X}}_{m}^{( t )T}+\mathbf{D}_{1}^{( t-1 )} , \ \ \ \   \mathbf{D}_{1}^{( t-1 )}=\widetilde{\mathbf{B}}^{( t )} \widetilde{\mathbf{X}}_{m}^{( t )T}, \\
& \mathbf{D}_{2}^{( t )}=\overrightarrow{\mathbf{X}}_{m}^{( t )} \overrightarrow{\mathbf{X}}_{m}^{( t )T}+\mathbf{D}_{2}^{( t-1 )} , \ \ \ \   \mathbf{D}_{2}^{( t-1 )}=\widetilde{\mathbf{X}}_{m}^{( t )}  \widetilde{\mathbf{X}}_{m}^{( t )T}.
\end{aligned}
\end{equation}
In above equations of $\mathbf{D}_{1}^{( t )}$ and $\mathbf{D}_{2}^{( t )}$, we can easily observe that only the first term needs to be calculated at round $t$ and their second terms $\mathbf{D}_{1}^{( t-1 )}$ and $\mathbf{D}_{2}^{( t-1 )}$can be directly obtained from last round. Similarly, $\mathbf{D}_{1}^{( t )}$ and $\mathbf{D}_{2}^{( t )}$ can be saved for the use in the next round so that the optimization at $(t+1)$ round will be efficient. Moreover, in order to capture nonlinear characteristics, we employ kernel features $\phi(\textbf{X}_1)$ as replacements for the original image features. Specifically, we utilize the Radial Basis Function (RBF) kernel mapping, represented as $\phi(\textbf{x})=exp(\frac{\| \textbf{x}-\textbf{a}_{i}\| _{2}^{2}}{2\sigma^2})$, where ${\{\textbf{a}_i}\}_{i=1}^m$denotes a set of randomly selected anchor points from the training samples in the first round, and $\sigma$ represents the kernel width.

\subsubsection{Fine-Grained Weights Learning}
Enhancing performance by the utilization of complementary multi-modal features is a fundamental goal in multi-modal hashing. Existing multi-modal hashing methods manually adjust balancing parameters for different modalities, a practice that has been shown to be less effective, as demonstrated in \cite{lu2019flexible}. In contrast, to more effectively fuse multi-modal information, FOMH \cite{lu2019flexible} employs a sophisticated weighting strategy capable of learning modality weights rather than relying on manual selection. While the effectiveness of this approach has been demonstrated, it is important to note that the learned weights are coarse-grained. Specifically, FOMH assigns identical weights to modalities for all query instances, overlooking the specific characteristics of individual querying data samples. 

In this paper, we introduce finer-grained weights to facilitate the effective fusion of complementary multi-modal features by capturing detailed information from each querying instance. The term 'fine-grained' here denotes that we learn distinct modality weights for each specific instance, meaning that each querying instance has its own unique set of weights.

For a given instance, if its image feature is more suitable for hash code learning, our objective is to magnify the influence of the image modality on hash code generation. Conversely, if the text modality of the instance is better suited for hash code learning, we want to increase the importance of the text modality. Based on this rationale, we design a variable for quantifying the significance of different modalities for each instance during the hash code generation. To achieve this, we have devised the following loss function, which facilitates the learning of a mapping that accurately reflects the importance of various modal features for each instance.
\begin{equation}\label{Fine-grained Weights}
\begin{aligned}
&\underset{\mathbf{U}_{m}^{( t )}}{\min}\sum_{m=1}^M{( \| \overrightarrow{\mathbf{B}}^{( t )}-\mathbf{W}_{m}^{( t )} \overrightarrow{\mathbf{X}}_{m}^{( t )} -\mathbf{U}_{m}^{( t ) T} \overrightarrow{\mathbf{X}}_{m}^{( t )} \| _{F}^{2} } \\ &{ +\| \widetilde{\mathbf{B}}^{( t )}-\mathbf{W}_{m}^{( t )} \widetilde{\mathbf{X}}_{m}^{( t )} -\mathbf{U}_{m}^{( t ) T}\widetilde{\mathbf{X}}_{m}^{( t )} \| _{F}^{2}+\delta \| \mathbf{U}_{m}^{( t )} \| _{F}^{2} )},
\\
\end{aligned}
\end{equation}
where ${\mathbf{U}}_{m}^{(t)}\in \mathbb{R} ^{d_{m}\times r}$ is the auxiliary projection to calculate fine-grained weights and $\delta$ is the parameter controls regularization term. In Eqn.(\ref{Fine-grained Weights}), $\mathbf{U}_{m}^{( t )}$  is learned from both new and old data so as to ensure the knowledge obtained in the past still contributes to the learning at the current round. 
To be noted that Eqn.(\ref{Hash Function Loss}) is designed with a structural risk minimization strategy to optimize the hash function, which is also verified to be a necessary term, as in Table  (\ref{ablation_table}). Different from the empirical risk minimization, this loss may naturally keep gaps between  $\overrightarrow{\mathbf{B}}^{( t )}$ and $\mathbf{W}_{m}^{( t )}\overrightarrow{\mathbf{X}}_{m}^{( t )}$ to obtain good generalization ability. Intuitively, we propose Eqn.(\ref{Fine-grained Weights}) to reflect the quality of the feature for learning hash codes. When generating hash codes for various queries, fine-grained weights could be obtained and wisely combined with multi-modal features for multi-modal hash code learning.

For optimizing Eqn.(\ref{Fine-grained Weights}), we can take the derivative w.r.t. $\mathbf{U}_{m}^{( t )}$ to zero, and the solution of $\mathbf{U}_{m}^{( t )}$ can be calculated as,
\begin{equation}\label{U}
\begin{aligned}
&\mathbf{U}_{m}^{( t )}=( \mathbf{D}_{2}^{( t )}+\theta \mathbf{I} ) ^{-1}( \mathbf{D}_{3}^{( t )}-\mathbf{D}_{2}^{( t )}\mathbf{W}_{m}^{( t ) T} ),
\end{aligned}
\end{equation}
where   
\begin{equation}
\begin{aligned}
\mathbf{D}_{3}^{( t )}=\overrightarrow{\mathbf{X}}_{m}^{( t )} {\overrightarrow{\mathbf{B}}^{( t )T}}+\mathbf{D}_{3}^{( t-1 )}, \ \ \  \mathbf{D}_{3}^{( t-1 )}=\widetilde{\mathbf{X}}_{m}^{( t )} \widetilde{\mathbf{B}}^{( t ) T}.
\end{aligned}
\end{equation}

When out-of-sample queries $\mathbf{X}_{q}$ come in querying period, we first generate the fine-grained weight $\mathbf{z}_{m}^{\left( t \right)}$  with the help of the learned $\mathbf{U}_{m}^{( t )}$. The specific formula is,
\begin{equation}
\begin{aligned}
\mathbf{z}_{m}^{\left( t \right)}=h_{\max}\mathbf{1}^T-\mathbf{h}_{m}^{\left( t \right)}, m=1,2,
\end{aligned}
\end{equation}
where
\begin{equation}
\begin{aligned}
&\mathbf{h}_{m}^{\left( t \right)}=\mathbf{1}^T\left| \mathbf{U}_{m}^{\left( t \right) T}  \mathbf{X}_{qm} \right|, m=1,2,
\\
&h_{\max}=\underset{m \in \left\{ 1,2 \right\},\,\,j \in \left\{ 1,2...n_q \right\}}{\max}\left( \mathbf{h}_{mj}^{\left( t \right)} \right),
\end{aligned}
\end{equation}
where $\mathbf{1}$ is an all-one vector, $\mathbf{X}_{qm}$ is the $m$-th modality feature, $n_q$ is the number of query data, $\left|\cdot \right|$ calculates absolute values for each element, and $\mathbf{z}_{m}^{\left( t \right)} (m=1,2)$ are the \textbf{fine-grained weights}. In above equations, $\mathbf{h}_{m}^{\left( t \right)}$ measures the quantization error (the gap between $\overrightarrow{\mathbf{B}}^{( t )}$ and $\mathbf{W}_{m}^{( t )}\overrightarrow{\mathbf{X}}_{m}^{( t )}$) for each instance and $\mathbf{z}_{m}^{\left( t \right)}$ transforms the error into weights by the maximum and subtract computations.

Then, the hash code of query can be realized by fusing heterogeneous modalities as,
\begin{equation}\label{hash combine}
\begin{aligned}
&\mathbf{B}_q=\mathrm{sign}(\sum_{m=1}^2{( \mathbf{1z}_{m}^{( t )} ) \odot \mathbf{W}_{m}^{( t )}\mathbf{X}_{qm}} ).
\end{aligned}
\end{equation}
where $\mathrm{sign}( \cdot ) $ is the sign function and $\odot$ is Hadamard product.

\subsection{Model Analysis}
We analyze the time complexity of our optimization. If there are new categories at the $t$-th round, we should generate high-level hash codes for them. The time complexity of generating $\overrightarrow{\mathbf{Y}}^{( t )}$ with Eqn.(\ref{high-level semantics}) is relevant to $c_{n}^{\left( t \right)}$. And Learn $\overrightarrow{\mathbf{B}}_{cj}^{\left( t \right)}$ using Eqn.(\ref{B_cj}) costs $O\left( \left( \left( r-1 \right) +rf+2 \right) c_n \right)$. Since $\overrightarrow{\mathbf{B}}_{c}^{\left( t \right)}$ has $r$ rows, updating $\overrightarrow{\mathbf{B}}_{c}^{\left( t \right)}$  totally costs $O\left( \left( \left( r-1 \right)+rf+2 \right) c_nr \right)$. Moreover, Learing $\mathbf{W}_c$ according to Eqn.(\ref{W_c}) costs $O\left( \left( r+  r^2+f+rf \right) \left( c_n+c_o \right)+r^2\left( r+f \right) \right)$. Because the number of max iterations in row three is a tiny constant and we take $5$ in our experiment, the whole time complexity of the training of high-level hash codes of categories is not related to $n^{(t)}$ and it is needed only if new categories appear at current round. Furthermore, generating the hash codes of instances with Eqn.(\ref{B}) spends $O( r\left( c_n+c_o \right) n^{\left( t \right)} ) $. The time complexity of learning $\mathbf{W}_{m}^{( t )}$ according to Eqn.(\ref{W}) is $O( ( 2n^{\left( t \right)}+rn^{\left( t \right)}+r+d_mn^{\left( t \right)}+2d_m+rd_m+d_{m}^{2} ) d_m ) $, and the time complexity of $\mathbf{U}_{m}^{( t )}$ is $O(( 2n^{( t )}+2d_mn^{( t )}+3d_m+d_{m}^{2}+ {3r}+{2d_mr}+n^{( t )}r ) d_m+rn^{( t )})$. From the above analysis, we could conclude that our approach is scalable for large-scale online multi-modal retrieval as its complexity is only linear with the size of newly coming samples $n^{(t)}$.

\section{Experiment}
\subsection{Experimental Settings}

\subsubsection{Datasets and Evaluation Metric}
In our experimental evaluation, we employed two widely used benchmark datasets. The \textbf{MIRFlickr} \cite{huiskes2008mir} dataset comprises $25,000$ instances distributed across $24$ categories. Following the preprocessing steps in \cite{jiang2019discrete}, which removes instances with tags appearing fewer than 20 times, $20,015$ image-text pairs are left. For feature extraction, we utilized the VGG network to generate $4096$-dimensional deep features for the image modality and employed $1386$-dimensional Bag of Words (BOW) features to represent the text modality. The \textbf{NUS-WIDE} \cite{chua2009nus} dataset is composed of $269,648$ instances spanning $81$ categories. Following the approach described in \cite{lu2019flexible}, we kept the top $21$ categories with the most occurrences, leaving $195,834$ instances at the end. Similar to our processing of the MIRFlickr dataset, we used the VGG network to generate $4096$-dimensional deep features from the original images and employed the $5018$-dimensional BOW features provided by the original dataset to represent the text features. In addition, both MIRFlickr and NUS-WIDE are multi-label datasets, where two instances are considered to have a ground-truth similarity if they share at least one common label. 

For evaluation, we followed the common practice in online multi-modal hashing \cite{xie2017dynamic,lu2019flexible} and utilized the Mean Average Precision (MAP) as our evaluation metric. Larger MAP values indicate superior performance. Please note that in online hashing literature \cite{lu2019flexible,weng2020online,wang2020online,wang2020label,yi2021efficient}, using only MAP is one of the standard settings.

\subsubsection{Baselines and Implementation Details}
In this study, we utilized several state-of-the-art multi-modal hashing methods as baselines, namely, OMH-DQ \cite{lu2019online}, SAPMH \cite{zheng2020adaptive}, FOMH \cite{lu2019flexible}, and OASIS \cite{wu2021online}. It's worth noting that FOMH and OASIS are online models, while the others operate in a batch-based manner. We obtained comparison results using the publicly available source code for these baseline methods. For batch-based models, they are trained on the entire dataset accumulated up to the current round to adapt to the online setting.

In our experiments, we set the parameters $\theta =1$ and $\delta =1 $. For class-level hash code learning, we employed $5$ iterations. All experiments are conducted on a Linux workstation equipped with an Intel XEON E5-2650 2.20GHz CPU and 128GB of RAM. To ensure robustness, we performed experiments multiple times and reported the average results.

\subsubsection{Online Experimental Settings}
In this study, we considered three distinct online settings to evaluate our proposed method.
\textbf{Standard Online Scenario (IID)}: This scenario, also known as the Independent Identically Distributed scenario, is a standard setting in online hashing literature. It assumes that no new categories will emerge with incoming data streams.
\textbf{Category Incremental Scenario (Overlap)}: This scenario is specifically designed for multi-label data, which is the case for both benchmark datasets, MIRFlickr and NUS-WIDE. Here, we assume that new data chunks may contain both previously encountered categories and new ones. 
\textbf{Category Incremental Scenario (Non-overlap)}: This scenario is frequently used \cite{RebuffiKSL17, LiH18a, WuCWYLGF19}, particularly for single-label datasets such as CIFAR \cite{Learning}. In this setting, categories in different chunks do not overlap.

For the sake of a fair comparison, we adopted the same experimental settings as those in OASIS \cite{wu2021online}. These settings encompass three scenarios. \emph{The Standard Online Scenario (IID)}: In this scenario, we randomly selected $2,000$ samples to form the test set, leaving the remaining samples for the training set. For the MIRFlickr dataset, we randomly divided it into $10$ chunks, with the first $9$ chunks each containing $2,000$ samples, and the last chunk consisting of $15$ samples. Similarly, for NUS-WIDE, we divided it into $20$ rounds, with the first $19$ chunks each containing $10,000$ samples and the last chunk comprising the remaining $3,834$ samples.
\emph{The Category Incremental Scenario (Overlap):} Both MIRFlickr and NUS-WIDE datasets are divided into $10$ and $20$ chunks, respectively. In this scenario, labels are allocated for each round, ensuring that the labels for new rounds include those from previous rounds, along with at least one new category. Subsequently, data chunks are constructed according to their assigned label sets. For each round, the training and test data were randomly selected within their respective data chunks at a $9:1$ ratio.
\emph{The Category Incremental Scenario (Non-overlap):} In this scenario, MIRFlickr is divided into $4$ data chunks, while NUS-WIDE is divided into $8$ chunks. Labels are assigned for each round, with an emphasis on ensuring that the labels for one round are entirely distinct from those in other rounds. Subsequently, samples are selected to form data chunks based on the assigned labels. As in the other scenarios, the training and test data for each round are randomly selected within their respective data chunks at a $9:1$ ratio.

Since some baselines are batch-based, those models utilized the entire training dataset accumulated from the first round to the current round as the database in all these scenarios.

\begin{table*}[!t] 
\scriptsize
\renewcommand{\arraystretch}{1.1}
\centering
\caption{MAP results of last round in standard online scenario on MIRFlickr. The best results are shown in black bold font.}
\begin{tabular}{cccccc}\toprule
{Model types} & {Models}& 32-bit  & 64-bit & 96-bit    & 128-bit \\\midrule
\multirow{2}{*}{Batch-based}
& SAPMH\cite{zheng2020adaptive}   &0.7039$\pm$0.0109&  0.7111$\pm$0.0017&  0.7133$\pm$0.0038&  0.7117$\pm$0.0028 \\
& OMH-DQ\cite{lu2019online}   &  0.6740$\pm$0.0162&  0.6878$\pm$0.0099&  0.6859$\pm$0.0095&  0.6932$\pm$0.0012   \\ \hdashline
\multirow{3}{*}{Online} 
& FOMH\cite{lu2019flexible}   &  0.5930$\pm$0.0127&  0.5965$\pm$0.0174&  0.5990$\pm$0.0154&  0.5858$\pm$0.0321\\
& OASIS\cite{wu2021online}   & \textbf{0.8557$\pm$0.0022}  &  0.8622$\pm$0.0021& 0.8637$\pm$0.0010 &  0.8651$\pm$0.0014   \\
& HCFW  &  0.8477$\pm$0.0043&  \textbf{0.8679$\pm$0.0037} &  \textbf{0.8829$\pm$0.0009}&  \textbf{0.8928$\pm$0.0042}
\\
\bottomrule
\end{tabular}\label{map1_last_round_mir}
\end{table*}

\begin{table*}[!t] 
\scriptsize
\renewcommand{\arraystretch}{1.3}
\centering
\caption{MAP results of last round in standard online scenario on NUS-WIDE. The best results are shown in black bold font.}
\begin{tabular}{cccccc}\toprule
{Model types} & {Models}& 32-bit  & 64-bit & 96-bit    & 128-bit \\\midrule
\multirow{2}{*}{Batch-based}& SAPMH\cite{zheng2020adaptive}   &0.5565$\pm$0.0484&  0.5354$\pm$0.0700&  0.5652$\pm$0.0581&  0.5951$\pm$0.0182    \\
& OMH-DQ\cite{lu2019online}   &  0.5417$\pm$0.0149&  0.5703$\pm$0.0080&  0.5719$\pm$0.0032& 0.5785$\pm$0.0188    \\ \hdashline
\multirow{3}{*}{Online} & FOMH\cite{lu2019flexible} &  0.6019$\pm$0.0102&  0.6145$\pm$0.0206& 0.6181$\pm$0.0068 &0.6150$\pm$0.0156    \\
& OASIS\cite{wu2021online}   &  0.7939$\pm$0.0061&  0.8006$\pm$0.0061&  0.8025$\pm$0.0055& 0.8055$\pm$0.0021    \\
& HCFW&  \textbf{0.8241$\pm$0.0107}&  \textbf{0.8517$\pm$0.0087}&  \textbf{0.8608$\pm$0.0056}& \textbf{0.8668$\pm$0.0020}
\\
\bottomrule
\end{tabular}\label{map1_last_round_nus}
\end{table*}

\begin{figure*}[t]
\centering
\includegraphics[width=0.99\textwidth]{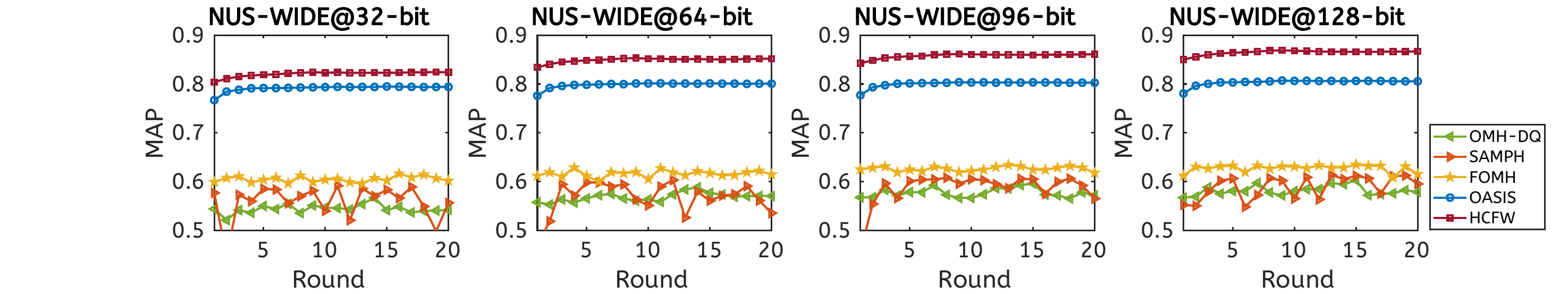}
\caption{ MAP results versus rounds in standard online scenario on NUSWIDE. }
\label{nus}
\end{figure*}

\subsection{Comparison with Baselines}
\subsubsection{MAP Comparisons in standard Online Scenario}
To assess the efficacy of our model, we initially conducted experiments in the standard online scenario. The retrieval results in terms of Mean Average Precision (MAP) for the final round are presented in Table \ref{map1_last_round_mir} and Table \ref{map1_last_round_nus}. Furthermore, we have plotted the MAP results across each round on the NUS-WIDE dataset in Figure \ref{nus}. From these results, we can observe that,
\begin{itemize}[leftmargin=*]
\item It can be seen that our proposed method consistently outperforms all selected baselines across the majority of cases. Several factors may account for this phenomenon: (1) 
Our approach introduces high-level hash codes, ensuring the consistency of hash codes during long-term learning. Regardless of which round the data, their hash codes are generated from invariant high-level hash codes, preserving semantic information throughout multiple rounds of training without loss. This strategy enhances the quality of online hash learning, which is also evident in the significant improvement observed during the long-term learning of the NUSWIDE dataset with 20 chunks. (2) Through the introduction of fine-grained weights, our method maximizes the utility of multi-modal heterogeneous features, leading to the generation of precise hash codes.
\item On the MIRFlickr dataset with 32-bit hash codes, our method exhibits slightly lower performance than OASIS \cite{wu2021online}. This can be attributed to the fact that when the number of hash bits is limited, high-level codes may not encapsulate sufficient semantic information. Nevertheless, considering the overall performance across most cases, our model remains effective.
\item Our method excels in retaining previously acquired knowledge throughout all stages of learning, resulting in a consistent upward trend in MAP results across rounds. Importantly, OASIS \cite{wu2021online} also exhibits effective retention of prior knowledge, reflecting a similar trend to ours. In contrast, FOMH \cite{lu2019flexible} displays slight instability in its MAP results across rounds, potentially attributable to its focus on preserving old knowledge primarily in hash function learning while neglecting it in hash code learning. Meanwhile, SAPMH \cite{zheng2020adaptive} and OMH-DQ \cite{lu2019online}, being batch-based methods, perform similarly at each round, as they are trained on the entirety of accumulated data.
\end{itemize}

\begin{table*}[!t]
\scriptsize
\renewcommand{\arraystretch}{1.3}
\centering
\caption{MAP results of last round in the category incremental scenario (overlap) and the category incremental scenario (non-overlap) on MIRFlickr. The best results are shown in black bold font.}
\begin{tabular}{cccccc}\toprule
{Settings}& {Models}& 32-bit  & 64-bit & 96-bit    & 128-bit \\\midrule
Category incremental &  OASIS\cite{wu2021online}  & 0.8058$\pm$0.0129& 0.8200$\pm$0.0091 &  0.8209$\pm$0.0051& 0.8227$\pm$0.0129  \\ 
scenario (overlap) & HCFW & 0.8025$\pm$0.0143 & 0.8281$\pm$0.0127 & 0.8431$\pm$0.0143 & 0.8497$\pm$0.0113 \\ \midrule
Category incremental&  OASIS\cite{wu2021online} & 0.3624$\pm$0.0960 & 0.3834$\pm$0.0526 & 0.4190$\pm$0.0470 & 0.3839$\pm$0.0951 \\
scenario (non-o verlap)& HCFW & 0.5228$\pm$0.0521 & 0.5420$\pm$0.0426 & 0.5671$\pm$0.0428 & 0.5565$\pm$0.0528
\\\bottomrule
\end{tabular}\label{map2_last_round_mir}
\end{table*}

\begin{table*}[!t]
\scriptsize
\renewcommand{\arraystretch}{1.3}
\centering
\caption{MAP results of last round in the category incremental scenario (overlap) and the category incremental scenario (non-overlap) on NUS-WIDE. The best results are shown in black bold font.}
\begin{tabular}{cccccc}\toprule
{Settings}& {Models}& 32-bit  & 64-bit & 96-bit    & 128-bit \\\midrule
Category incremental &  OASIS\cite{wu2021online}  & 0.7881$\pm$0.0085 & 0.7951$\pm$0.0093 & 0.8083$\pm$0.0103 & 0.7955$\pm$0.0115  \\ 
scenario (overlap) & HCFW  & 0.7908$\pm$0.0133 & 0.8103$\pm$0.0114 & 0.8415$\pm$0.0106 & 0.8429$\pm$0.0109 \\ \midrule
Category incremental&  OASIS\cite{wu2021online} & 0.2650$\pm$0.0537 & 0.2927$\pm$0.0632 & 0.2918$\pm$0.0891 & 0.2875$\pm$0.0439 \\
scenario (non-overlap)& HCFW& 0.3936$\pm$0.0623 & 0.4354$\pm$0.0482& 0.4273$\pm$0.0345 & 0.4339$\pm$0.0583
\\\bottomrule
\end{tabular}\label{map2_last_round_nus}
\end{table*}

\subsubsection{MAP Comparisons in Category Incremental Scenarios}
Additionally, we conducted tests in both the category incremental scenario (overlap) and the category incremental scenario (non-overlap) to highlight the advantages of our method. It's important to note that among the baselines, only OASIS \cite{wu2021online} possesses the capability to address the category incremental problem. Consequently, comparing our method with other baselines in this subsection would be unfair. Thus, we only compare our method with OASIS. The results for the last round's MAP are presented in Table \ref{map2_last_round_mir} and Table \ref{map2_last_round_nus}. Key observations include:
\begin{itemize}[leftmargin=*]
\item Notably, under the category incremental scenarios, our model exhibits a more substantial improvement compared to OASIS \cite{wu2021online} than in standard settings. This effect may be attributed to the approach of OASIS, which overlooks the fusion of multi-modal features when generating hash codes. In addition to this, HCFW introduces high-level hash codes that can explicitly ensure the compactness of hash codes for the same category across multiple rounds, as well as the similarity of hash codes for similar-category data across multiple rounds in online setting. Such a strategy enables HCFW to generate more accurate hash codes.
\item Our method exhibits slightly lower performance than OASIS \cite{wu2021online} with 32-bit hash codes. The reasons are the same as the above analysis, i.e., when the number of hash bits is limited, high-level codes may not encapsulate sufficient semantic information.
\end{itemize}

\subsection{Model Analysis}
\begin{table*}[!t]
\scriptsize
\renewcommand{\arraystretch}{0.9}
\centering
\caption{MAP results between OASIS with fine-grained weights and our method. The only difference between them is whether using the high-level codes, which can further evaluate the effectiveness of the high-level codes. FW means the fine-grained weights. The best results are shown in black bold font. We only show some rounds of MAP results because of the space limit.}
\begin{tabular}{c|c|c|ccccccccc}\toprule
Models & {FW}& {HC} & r1 & r3  & r5 & r10  & r13  & r15  & r20 \\\midrule
OASIS \cite{wu2021online} & & &0.7770& 0.7974& 0.8012 & 0.8030& 0.8032& 0.8033& 0.8025\\
OASIS + FW  & \checkmark& & 0.8069& 0.8123& 0.8141 & 0.8160& 0.8162& 0.8164& 0.8153\\
HCFW (Ours) & \checkmark&\checkmark&\textbf{0.8208} & \textbf{0.8454} & \textbf{0.8504} & \textbf{0.8533} & \textbf{0.8541} & \textbf{0.8548} & \textbf{0.8551}
\\\bottomrule
\end{tabular}\label{ablation:high_level}
\end{table*}

\subsubsection{On High-Level Codes}\label{model_design} In order to further evaluate the importance of the high-level codes, we conducted an ablation study, adding fine-grained weights into the state-of-the-art method OASIS to fairly compare with our method (the only difference is whether using the high-level codes). The experiments are conducted under the standard online setting on the NUS-WIDE dataset with the code length of 96 bits. The results are shown in Table \ref{ablation:high_level}. It can be seen that:
\begin{itemize}[leftmargin=*]
\item The high-level codes strategy significantly outperforms the hash code learning strategy in OASIS. The reason is that high-level code maintains consistency during long-term online learning, keeping invariant hash code generation in different rounds and ensuring the quality of the hash code.
\item The performance increase grows with the number of rounds, which further verifies our perspective that the high-level codes ensure consistency and quality during online learning. As the number of rounds increases, the hash codes of instances with the same or similar categories would become more and more inconsistent in OASIS, while our method can always keep the consistency during online learning.
\item It is surprising that adding fine-grained weights to OASIS can further improve the performance from $0.8025$ to $0.8153$. It further verifies the universality and robustness of the fine-grained weights, which can be used as a plug-and-play module for all the multi-modal hashing methods.
\end{itemize}
In addition, we conducted experiments using alternative forms of supervision and text-embedding strategies to learn high-level codes.

\begin{table}
\scriptsize
\centering
\caption{MAP results with and without fine-grained weights. FW means ``Fine-Grained Weights'' and RT means ``Regularization term ''. The best results are shown in black bold font. }
\begin{tabular}{c|c|c|cccccccc}\toprule
\multirow{2}{*}{Settings}& \multirow{2}{*}{FW}& \multirow{2}{*}{RT}&\multicolumn{4}{c}{MIRFlickr} 
\\\cmidrule(lr){4-7}
& & & 32-bit  & 64-bit & 96-bit    & 128-bit  \\\midrule
\multirow{3}{*}{IID}& & \checkmark& 0.8412 & 0.8657 & 0.8824 & 0.8907   \\
& \checkmark & & 0.8523 & 0.8718 & 0.8817 & 0.8792   \\
&\checkmark &\checkmark & \textbf{0.8606}& \textbf{0.8759}& \textbf{0.8893}& \textbf{0.8928}\\ \midrule
category& & \checkmark& 0.4892 & 0.5183& 0.5002& 0.5499\\
incremental&\checkmark & & 0.1744 & 0.1430 & 0.1288 & 0.1673   \\
(non-overlap)&\checkmark &\checkmark & \textbf{0.5062}& \textbf{0.5481}& \textbf{0.5514}& \textbf{0.5861} \\
\bottomrule
\end{tabular}\label{ablation_table}
\end{table}

\subsubsection{On Fine-Grained Weights }\label{ablation}
To empirically assess the significance of fine-grained weights, we conducted an ablation study by removing the fine-grained weights module from HCFW. The ablation study was performed under both the standard online setting and the category incremental scenario (non-overlap) on the MIRFlickr dataset. The results are presented in Table \ref{ablation_table}. The key findings are: Fine-grained weights exert a notable influence on the model's performance, particularly in the category incremental scenario. This enhancement is evident from the considerable performance gains achieved, underscoring the value of fine-grained weights.
The presence of fine-grained weights bridges the natural gap between $\overrightarrow{\mathbf{B}}^{( t )}$ and $\mathbf{W}_{m}^{( t )}\overrightarrow{\mathbf{X}}_{m}^{( t )}$, contributing significantly to overall performance improvement.

To further prove the necessity of the fine-grained weights, we performed an experiment on the MIRFlickr dataset to emphasize the necessity of incorporating a regularization term into hash function learning, which is a prerequisite for fine-grained weights usage. The results are presented in Table \ref{ablation_table}. Key observations include:
\begin{itemize}[leftmargin=*]
\item When regularization terms are omitted, there is a notable degradation in performance. This decline is particularly pronounced in category-incremental scenarios where sample sizes are limited. In such cases, the absence of regularization terms results in pronounced overfitting of the model.
\item The presence of a regularization term in hash function learning is crucial. It creates a natural gap between $\overrightarrow{\mathbf{B}}^{( t )}$ and $\mathbf{W}_{m}^{( t )}\overrightarrow{\mathbf{X}}_{m}^{( t )}$, and our fine-grained weights strategy becomes meaningful in this context.  Moreover, incorporating fine-grained weights also enhances retrieval performance.
\end{itemize}

\begin{figure}[t]
\centering
\includegraphics[width=0.97\columnwidth]{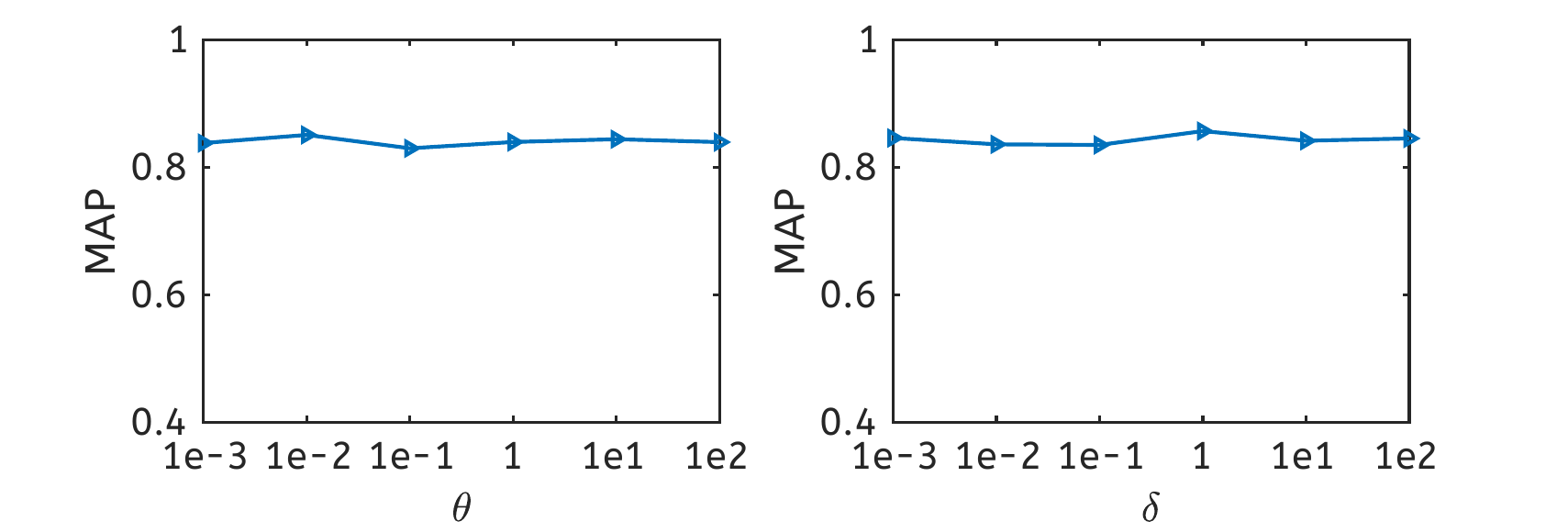}
\caption{Parameter sensitivity results on MIRFlickr.}
\label{sensitive}
\end{figure}

\subsubsection{On Parameter Sensitivity}
In our quest to understand the sensitivity of our model's performance to various parameter values, we conducted rigorous experiments under the standard online setting on the MIRFlickr dataset. We focused on examining the impact of parameter variations while holding other parameters constant. The outcomes of these experiments are presented in {Figure \ref{sensitive}}, with code length set at 32 bits. It can be seen that:
\begin{itemize}[leftmargin=*]
\item Notably, our model exhibits robustness to changes in most parameters. This robustness may be attributed to the ease with which our method's loss function converges during training. Consequently, our model demands minimal parameter tuning, rendering it highly promising for real-world applications.
\item In this paper, we empirically adopted the following parameter setting $\{\theta =1, \delta =1 \} $.
\end{itemize}
In addition, we conducted convergence experiments and time comparisons to demonstrate the effectiveness and robustness of our method HCFW.

\subsection{On Convergence}
\begin{figure}[t]
\centering
\includegraphics[width=0.99\columnwidth]{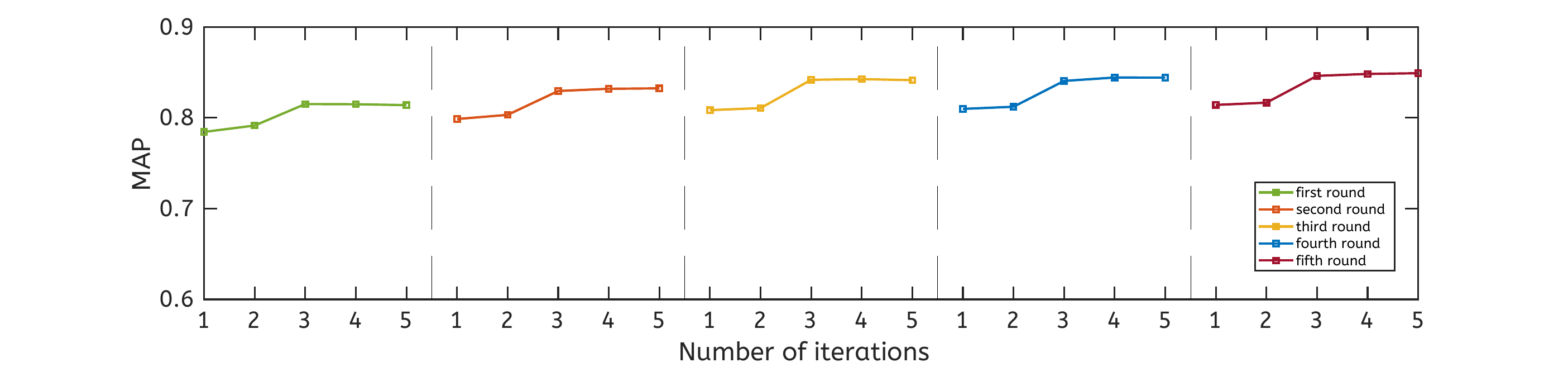}
\caption{Convergence analysis on MIRFlickr. }
\label{convergence}
\end{figure}

Furthermore, we validated the convergence of the alternative optimization of category-level hash code learning. Under the standard online setting, experimental results in the case of $32$ bits on MIRFlickr are shown in {Figure \ref{convergence}}. It can be found that our model could quickly converge. This phenomenon reflects that our optimization strategy is robust and the loss function is easy to learn. Considering both efficiency and performance, we set the number of iterations as $5$ in all experiments.

\begin{table*}[!t]
\scriptsize
\renewcommand{\arraystretch}{1.3}
\centering
\caption{Time results versus rounds.}
\begin{tabular}{cccccc}\toprule
Models & round 1 & round 2 & round 3  & round 4 &round 5 \\\midrule
SAPMH\cite{zheng2020adaptive} & 2.2193 & 2.7865 & 3.7431& 3.7686& 4.0627\\ 
OMH-DQ\cite{lu2019online} & 1.2987 & 1.3546 & 1.4379& 1.4834&1.6604 \\
FOMH\cite{lu2019flexible} & 0.3154 & 0.3632 & 0.3269& 0.4570& 0.2467 \\
OASIS\cite{wu2021online} & 2.4000 & 2.6239 & 2.6546& 2.6318& 2.5868 \\
HCFW (Ours) & 1.1345 & 1.1624 & 1.1435& 1.3864& 1.2055 \\
\hline\hline
Models & round 6  & round 7  & round 8 & round 9  & round 10 \\\midrule
SAPMH\cite{zheng2020adaptive} & 4.3185& 5.3329& 7.5435& 8.0551& 8.1584\\ 
OMH-DQ\cite{lu2019online}&  1.7603& 1.8037& 1.8329& 1.8221& 2.0590\\
FOMH\cite{lu2019flexible} & 0.4905& 0.3804& 0.3927& 0.3828&0.1089 \\
OASIS\cite{wu2021online} & 2.6082& 2.7016& 2.7943& 2.6911& 2.0635 \\
HCFW (Ours) & 1.1558& 1.0928& 1.1323& 1.2419& 1.2121
\\\bottomrule
\end{tabular}\label{time_table}
\end{table*}

\subsection{Time Comparisons}
To assess the computational efficiency of HCFW, we conducted time-cost comparisons across all methods in the standard online scenario using the MIRFlickr dataset, with hash codes of length 32 bits. The results, detailed in Table \ref{time_table}, reveal the following insights:
\begin{itemize}[leftmargin=*]
\item Batch-based hashing methods, such as SAPMH \cite{zheng2020adaptive} and OMH-DQ \cite{lu2019online}, exhibit substantially longer execution times compared to online hashing, emphasizing the imperative of addressing the online hashing problem.
\item FOMH demonstrates slightly faster training times than our HCFW. This discrepancy arises because FOMH's training process only constructs pairwise similarity within the new data chunk, whereas our method simultaneously incorporates both old and new data chunks.
\item Our proposed HCFW boasts rapid training, surpassing even the state-of-the-art category-incremental method, OASIS \cite{wu2021online}. This efficiency is attributed to our model's linear relationship between training time and new data, independent of the number of old data instances. Furthermore, when no new categories are introduced in a given round, our model's optimizations can be executed without the need for iterative processes. This strategic design significantly contributes to training efficiency.
\end{itemize}
In summary, our proposed HCFW demonstrates computational efficiency without compromising accuracy, positioning it as a practical choice for online multi-modal hashing, considering both speed and precision.

\begin{table*}[!t]
\renewcommand{\arraystretch}{1.3}
\centering
\scriptsize
\caption{MAP results of using different types of supervision strategies to learn high-level codes. The best results are shown in black bold font.}
\begin{tabular}{cccccc}\toprule
Supervision & round 1& round 2 & round 3 &round 4 &round 5 \\ \midrule
Hadamard & 0.8262& 0.8417& 0.8510& 0.8549& 0.8597 \\
Semantics & \textbf{0.8327}& \textbf{0.8486} & \textbf{0.8582} & \textbf{0.8612}& \textbf{0.8672} \\
\hline\hdashline
Supervision & round 6 & round 7 &round 8 & round 9& round 10\\ \midrule
Hadamard & 0.8617 &  0.8625& 0.8623& 0.8623&0.8624 \\
Semantics & \textbf{0.8699} & \textbf{0.8710} & \textbf{0.8704} & \textbf{0.8711} & \textbf{0.8710} \\
\bottomrule
\end{tabular}\label{model_design_supervision}
\end{table*}

\begin{table*}[!t]
\renewcommand{\arraystretch}{1.3}
\centering
\scriptsize
\caption{MAP results of using different types of text embedding strategies to learn high-level codes. The best results are shown in black bold font.}
\begin{tabular}{cccccc}\toprule
Text Embedding Strategy & round 1& round 2 & round 3 &round 4 &round 5 \\ \midrule
BERT\cite{devlin2018bert} & 0.8481& 0.8566& 0.8613& 0.8643& 0.8649 \\
Google word2vec\cite{mikolov2013efficient} & 0.8455&  0.8567& 0.8639& 0.8680& 0.8694 \\
Clip\cite{RadfordKHRGASAM21} & \textbf{0.8527}&  \textbf{0.8628}& \textbf{0.8695}& \textbf{0.8730}& \textbf{0.8755} \\
\hline\hdashline
Text Embedding Strategy & round 6 & round 7 &round 8 & round 9& round 10\\ \midrule
BERT\cite{devlin2018bert} & 0.8667& 0.8684& 0.8686 & 0.8693& 0.8693 \\
Google word2vec\cite{mikolov2013efficient} &  0.8704 & 0.8724 & 0.8733 & 0.8734& 0.8733 \\
Clip\cite{RadfordKHRGASAM21}&  \textbf{0.8764} & \textbf{0.8778} & \textbf{0.8786} & \textbf{0.8793}& \textbf{0.8794} \\
\bottomrule
\end{tabular}\label{model_design_embedding}
\end{table*}

\subsection{Model Design Experiments}
To gain a deeper understanding of our high-level codes module, we conducted experiments using alternative forms of supervision to learn high-level codes. Specifically, we employed the Hadamard matrix as supervision, replacing semantics for comparative analysis. The Hadamard matrix is widely recognized in online hashing literature and has demonstrated effectiveness \cite{li2022recent}. These experiments were conducted on MIRFlickr under standard online settings with a code length of 64 bits. The results are presented in Table \ref{model_design_supervision} From the table, several observations can be made:
\begin{itemize}[leftmargin=*]
\item Using different types of supervision to learn high-level codes yields comparable performance. These results underscore the generality and effectiveness of our strategy, leaving room for exploring additional supervision methods in future research.
\item The use of semantics as supervision outperforms the Hadamard matrix approach. This is attributed to semantics enabling the construction of semantic relevance between categories and their subsequent embedding into hash codes, providing a more interpretable solution.
\end{itemize}

Additionally, we assessed various types of text embedding strategies, including Word2Vec\cite{mikolov2013efficient}, BERT\cite{devlin2018bert}, and CLIP\cite{RadfordKHRGASAM21}. Experiments are conducted on MIRFlickr with a 64-bit code length under standard online settings. The results are presented in Table \ref{model_design_embedding}. From the table, it can be observed that these strategies yield relatively comparable performance, with the CLIP strategy exhibiting superior performance. However, it's important to note that the selection of text embedding strategy is not the primary focus of this paper. In all our experiments, we adopted the Google Word2Vec strategy \cite{mikolov2013efficient}, with a word vector dimensionality, denoted as $k$, set to 300.

\section{Conclusion}
In this paper, we introduce a novel approach termed High-level Codes, Fine-grained Weights (HCFW). To ensure long-term consistency in category-incremental scenarios, we introduce high-level codes as a solution. Furthermore, we devise a loss function for learning the hash codes of new categories by embedding semantic information. Additionally, we propose a novel strategy, referred to as fine-grained weights, to maximize the utilization of heterogeneous modalities and enhance hash learning by fusing multi-modal features within each instance. Thanks to its innovative model design, HCFW can be efficiently trained, delivering remarkable performance on benchmark datasets. Furthermore, we have taken strides to ensure the accessibility and reproducibility of HCFW by making its code and features publicly available.

\section{Acknowledgment}
This work was supported in part by the National Natural Science Foundation of China under Grant 62202278, 62172256, in part by Natural Science Foundation of Shandong Province under Grant ZR2019ZD06, in part by the Young Scholars Program of Shandong University, and in part by Open Foundation of Yunnan Key Laboratory of Software Engineering under Grant No.2023SE302. Special thanks to Professor Mohan Kankanhalli for his invaluable guidance and support throughout this research.

 \bibliographystyle{elsarticle-num} 
 \bibliography{HCFW}







\end{document}